\documentstyle[12pt]{article}
\def\n{\noindent}
\def\cl{\centerline}
\def\o{\over}
\def\p{\partial}
\def\be{\begin{equation}}
\def\ee{\end{equation}}

\def\eg{{e.g.}\,}

\def\ie{{i.e.}\,}
\def\etal{{et al.}\,}

\def\deriv#1#2{{{\rm d}#1\over {\rm d}#2}}            	% derivative
\def\pd#1#2{{\partial #1 \over \partial #2}}		% partial derivative
\def\gapprox{\;\rlap{\lower 2.5pt                     	% approx. greater
             \hbox{$\sim$}}\raise 1.5pt\hbox{$>$}\;}
\def\lapprox{\;\rlap{\lower 2.5pt                     	% approx. smaller
             \hbox{$\sim$}}\raise 1.5pt\hbox{$<$}\;}

\def\msun{M_{\odot}}

\def\mdot{\dot M}
\def\mdote{\dot M_{\rm Edd}}

\def\ref{\noindent \hangindent 20pt }

\def\aaa#1{{A\&A,} {#1}}

\def\annrev#1{{ARA\&A,} {#1}}

\def\apj#1{{ApJ,} {#1}}

\def\mnras#1{{MNRAS,} {#1}}
\def\nature#1{{Nature,} {#1}}
\def\pasj#1{{PASJ,} {#1}}

\def\ssr#1{{Space\ Sci.\ Rev.,\/} {#1}}
\begin{document}
\baselineskip=14pt

\cl{\bf LOW FREQUENCY QUASI-PERIODIC OSCILLATIONS IN LOW}
\cl{\bf MASS X-RAY BINARIES AND GALACTIC BLACK HOLE CANDIDATES}
\bigskip

\cl {\bf Xingming Chen and Ronald E. Taam}
\smallskip

\cl {Department of Physics and Astronomy}
\cl {Northwestern University, Evanston, IL 60208}
\medskip

\cl{To Appear in ApJ}

\vskip 1.0in

\cl{\bf ABSTRACT}
\medskip
\par
We consider the inner regions of accretion disks surrounding black
holes and neutron stars and investigate the nonlinear time dependent evolution
of thermal-viscous instabilities. The viscous stress is assumed to be
proportional to the gas pressure with the viscosity parameter formulated as
$\alpha =\min[\alpha_0 (h/r)^n, \alpha_{\max}]$,
where $h$ is the local scale height, $r$ is the distance from the central
compact object, and $n$, $\alpha_0$ and $\alpha_{\max}$ are constants.
It is found that the disk is unstable for $\alpha$
sufficiently sensitive to $h$ ($n \gapprox 1.2$). The instabilities are
globally coherent in the
entire unstable region of the disk and, depending on the viscosity parameters,
the time variability of the mass accretion rates are manifested as periodic or
quasi-periodic
oscillations. We show that, the low frequency ($\sim 0.04$~Hz) quasi-periodic
oscillations (QPOs) discovered recently in some of the black hole
candidates (Cyg~X-1 and GRO~J0422+32) and a low mass X-ray binary (Rapid
Burster MXB~1730--335) may be explicable by the
thermal-viscous instabilities in accretion disks. The observations of QPOs
place constraints on the viscosity parameters and suggest
that $(n,\alpha_0)$ $\sim (1.6,30)$ for the Rapid Burster with a $1.4\,\msun$
neutron star. In the case of black hole candidates, the dependence of
$\alpha$ on $h/r$ is less steep corresponding to $n \sim 1.2-1.3$ for black
holes less than $10\,\msun$.

\bigskip
\n{\it Subject headings:}

accretion, accretion disks --- instabilities ---
black hole physics --- stars: neutron

\vfil\eject

\cl{\bf 1. INTRODUCTION}
\medskip
\par
The quasi-periodic oscillation (QPO) phenomenon in low mass X-ray binaries
(LMXBs) and galactic black hole candidates (GBHCs)
provides a powerful tool to probe the temporal behavior of these
systems. An understanding of the observed luminosity variations may place
important constraints on the physics of accretion in these objects.
QPO phenomenon in bright LMXBs have been known since the
mid 1980s (see Lewin, van Paradijs, \& van der Klis 1988;
van der Klis 1989).
Two distinct kinds of QPOs have been classified, one is the intensity-dependent
QPOs characterized by frequencies $\sim 20 - 50$ Hz, and the other is the
intensity-independent QPOs with frequencies of $\sim 5-10$ Hz. The former QPOs
are called horizontal-branch QPOs
whereas the latter QPOs are called normal branch QPOs. The
horizontal-branch QPO phenomenon has been interpreted in terms of an
oscillatory variation in the luminosity due to modulations in the mass
accretion rate and the QPO frequencies are identified with a beat frequency
corresponding to
the difference between the Keplerian orbital frequency at the magnetopause
and the rotation frequency of the central neutron star (Alpar \& Shaham 1985;
Lamb, Shibazaki, Alpar, \& Shaham 1985).  The normal branch QPOs,
on the other hand, have been
modeled as a radiation instability in a spherical accretion flow (Fortner,
Lamb, \& Miller 1989) or as a variation in the vertical structure of an
accretion disk (Alpar, Hasinger, Shaham, \& Yancopoulos 1992) for accretion
rates near the Eddington limit. In contrast to the model for the horizontal
branch QPOs, the normal branch QPOs are interpreted in terms of optical depth
variations of a nearly constant luminosity.

More recently, another kind of QPOs from X-ray binary sources have been
discovered. They are characterized by low frequencies of $\sim$ 0.04 Hz
in both neutron star and black hole candidate systems.
For example, such QPOs have been found in the hump phase of
the persistent emission of the Rapid Burster MXB~1730--335 (Lubin et al. 1992,
1993), in the low state of Cyg~X-1 (Angelini \& White 1992; Vikhlinin \etal
1992a, 1994; Kouveliotou \etal 1992a), and in GRO~J0422+32 (Vikhlinin \etal
1992b, Kouveliotou \etal 1992b, Pietsch \etal 1993). While the Rapid Burster is
believed to be a neutron star, the other two sources are
black hole candidates. If a black hole is involved, the
mechanism deemed responsible for the production of these low frequency QPOs may
be different from those proposed for the other classes of QPOs
(the above models for the horizontal or normal branch QPOs require the central
objects to be neutron stars) since such oscillations would then be
independent of the nature of the compact object.
Thus, it is natural to explore the possibility
that disk instabilities are involved in the phenomenon.  Among the various
models proposed for the general QPO phenomenon is a model suggested by
Abramowicz, Szuszkiewicz, \& Wallinder (1989) who proposed that thermal and
viscous instabilities in accretion disks could be important for QPOs with
frequencies of $\sim 1$ Hz for LMXBs and $\sim 0.1$ Hz for GBHCs.

It is well known that the inner regions of viscous accretion disks surrounding
black holes or neutron
stars may suffer thermal-viscous instabilities when radiation pressure is
important (see \eg, Piran 1978). For a standard $\alpha$-model of accretion
disks (see Shakura \& Sunyaev 1973 ) with the viscous
stress, $\tau$, proportional to the total pressure, $p$, \ie,
$\tau = -\alpha p$, where $\alpha$ is a constant, disks are thermally and
viscously unstable if the pressue is radiation dominated
(Lightman \& Eardley 1974; Shakura \& Sunyaev 1976). However, if the viscous
stress is
proportional to the gas pressure, $p_g$, only, \ie, $\tau = -\alpha p_g$,
and $\alpha$ is a constant (see, \eg, Lightman \& Eardley 1974; Coroniti 1981;
Stella \& Rosner 1984), then the disk is always secularly stable.

Time dependent calculations of the thermal-viscous instabilities have been
carried out to examine the global behavior of the disk (see, \eg, Taam \&
Lin 1984; Honma,
Matsumoto \& Kato 1991; Lasota \& Pelat 1991). It was revealed that,
for the $\tau = -\alpha p$ prescription, the disk is globally unstable and
large-amplitude, burst-like fluctuations of the disk luminosity arise if the
mass accretion rates are high enough that the pressure in the disk is radiation
dominated. However, this strong instability may be reduced or even suppressed
by modifying the $\alpha$-viscosity prescription. For example, with
$\tau = -\alpha p \beta^m $ (see Abramowicz, Szuszkiewicz \& Wallinder 1989;
Szuszkiewicz 1990), where $\beta=p_g/p$,
the disk is stable for $m=0.5$ (Taam \& Lin 1984; Honma \etal 1991).
For $m=0.25$, the amplitude and the time scale of the fluctuations of the disk
luminosity is largely reduced (Honma \etal 1991). The stabilization in the
case of $m=0.5$ or effectively $\tau = -\alpha \sqrt {pp_g}$ in Keplerian
disks, is not predicted by local linear analysis. It is due to the advection of
energy by material motion in the radial direction (Taam \& Lin 1984).

Notwithstanding the lack of a fundamental theory for the viscosity in accretion
disks, the two phenomenological viscosity prescriptions, \ie,
$\tau = -\alpha p$ and $\tau = -\alpha p_g$, have some theoretical support
based on a turbulent viscosity picture (see Shakura \& Sunyaev 1973)
and magnetic field induced viscosity picture (see, \eg, Lightman \& Eardley
1974; Coroniti 1981; Stella \& Rosner 1984), respectively.
On the other hand, the disk instabilities based upon a viscous stress
proportional to
the total pressure may be too strong to be relevant to the low frequency QPOs,
and disks with the gas pressure prescription are stable and can not introduce
any variability within the assumed framework.
A possible resolution to this dilemma is to introduce a mixed prescription.
In other words,
as mentioned above, one may introduce a dimensionless parameter $m$ between
0 and 1. However, there is another possibility, namely, that the viscous
parameter, $\alpha$, is not a constant. In fact, a non-constant $\alpha$ is
usually required in the studies of accretion disks in dwarf novae.
In that case, the disk temperatures are
low and radiation pressue is unimportant in comparision with the gas
pressure. However, due to the sensitivity of the opacity on the temperature,
a thermal instability occurs. These instabilities have been applied to explain
the dwarf nova phenomena, and in order to understand both the
recurrence time and the duration of the outbursts, different values of the
viscosity parameter $\alpha$ are required during the two states.
One widely applied formula which can approximate such a variation is the form
of
$\alpha = \alpha_0 (h/r)^n$ (\eg, Meyer \& Meyer-Hofmeister 1983;
Duschl \& Livio 1989), where $h$ is the local scale height of the disk and $r$
is the distance from the compact object. This
form also has a theoretical basis from scaling arguments for
magnetic field induced viscosity. In particular, for a mean helicity determined
by cyclonic convective motions, Meyer \& Meyer-Hofmeister (1983) find
$n=1.5$, whereas for a helicity due to internal waves,
Vishniac \& Diamond (1992) conclude that $n=4/3$.
In addition, it was shown that, the effective $\alpha$ due to spiral shock
waves varies like $(h/r)^{1.5}$, even though its magnitude is very small
(see Spruit 1987).

The $(h/r)^n$ dependence has a destabilizing effect (assuming $n$ is
positive) on the disk if it is applied to the high temperature inner regions.
Based on the previous nonlinear studies, the $\tau=-\alpha p$ model is
excluded. We apply this form for $\alpha$ to the most
stable model, \ie, $\tau=-\alpha p_g$. In this case, it
has been shown that the disk is locally unstable (Meyer 1986) for $n
\gapprox 0.75$ in the limit that radiation pressure in dominant.
In a recent study, Milsom, Chen, \& Taam (1994, hereafter
MCT) generalized the results of Meyer (1986) and, furthermore, pointed out
that, in this model, unlike the $\tau=-\alpha p$ prescription, the instability
may take a milder form and non burst-like oscillations may exist in some
circumstances, which may be applied to an explanation of the low frequency
QPOs.

In this paper, we investigate the global evolution of the
thermal-viscous instabilities in accretion disks in a time dependent approach
to determine their viability as a mechanism for low frequency QPOs.
It is the purpose of this paper to report on the detailed nonlinear time
dependent calculations of accretion disks which suffer thermal-viscous
instabilities based upon the above modified form of the viscosity prescription.
In the next section, the fundamental equations and the approach we adopt are
outlined. In \S3 a brief review of the local thermal and viscous instability in
accretion disks is presented. The detailed numerical results of the global
evolution are given in \S4, and the implications of our results and
their possible applications for interpretation of observations will be
discussed in the last section.

\bigskip

\cl{\bf 2. FORMULATION}
\medskip
\par
We focus only on the thermal-viscous instability which may arise in accretion
disks. The axisymmetric inertial-acoustic instabilities, which have a much
shorter time scale and may exist possibly in the innermost regions of the disk
(see, \eg, Matsumoto, Kato, \& Honma 1989; Nowak \& Wagoner 1991;
Wallinder 1991; Chen \& Taam 1993), are not
considered here. Non-axisymmetric or self-gravitating effects (see,
\eg, Papaloizou \& Pringle 1984, 1985, 1987), are also neglected.
In other words, we assume a Keplerian disk, which is axisymmetric,
non self-gravitating, optically thick and geometrical thin so that it can be
described by the vertically integrated equations.
In this approximation, the surface density of the
disk, $\Sigma$, at a given cylindrical radius, $r$, can be obtained by
combining the conservation equations of mass and angular momentum. It is
given by the
standard time dependent mass diffusion equation (Lynden-Bell \& Pringle
1974) as
$$\pd \Sigma t =  {3 \o r} {\p \o \p r}\left[r^{1/2} {\p \o \p r}\left(
\nu \Sigma  r^{1/2} \right)\right],\eqno(1)$$
where $t$ is the time, and $\nu$ is the effective kinematic viscosity.
In equation (1), the Newtonian Keplerian angular velocity,
$\Omega = \sqrt{GM/r^3}$, is
implicitly assumed, where $M$ is the mass of the central compact object.
Relativistic effects on the gravitational potential are important
near the inner edge of the disk, but we anticipate that the long time scale
($\sim 20$ s) associated with the low frequency QPOs dictates that the
region of instability lies at radii sufficiently large compared to the
innermost disk that the results we obtain will be insensitive to the neglect
of these effects.

The mid-plane temperature, $T$, of the disk is governed by energy conservation
which is expressed as
$$C_v \Sigma T \left[\left({\pd {\ln T} t} +V_r {\pd {\ln T} r}\right)
-(\Gamma_3 -1)\left( {\pd {\ln \Sigma} t}+V_r{\pd {\ln \Sigma} r}
-{\pd {\ln H} t} \right)\right]$$
$$ = F^+ - F^- -{2\o r}{\partial {(rF_rH)} \o \partial r},\eqno(2)$$
where $V_r$, $\Gamma_3$, and $C_v$ are the radial velocity, the
adiabatic exponent and the specific heat at constant volume respectively.
$H$ is the half-thickness of the disk and is defined as
$$H =  {2p\o \Sigma\Omega^2}. \eqno(3)$$
The terms on the right hand side of equation (2) describe the local heating
and cooling processes,  where $F^+$ is the viscous dissipation rate, $F^-$ is
the cooling rate in the vertical direction, and $F_r$ is the radiative energy
flux in the radial direction. The heating rate and viscosity is related by
$$F^+ = \nu\Sigma {\left(r{\pd \Omega r}\right)}^2.\eqno(4)$$
And the energy transport flux in the radial direction is written as
$$F_r =- 2H F^- \pd {\ln T} r.\eqno(5)$$

The effective kinematic viscosity is parameterized in terms of the $\alpha$
model and is assumed to be
$$\nu={2 \o 3} \alpha c_s h (p_g/p), \eqno(6)$$
where $c_s$ and $h$ are the local sound speed and scale height respectively.
The viscosity parameter, $\alpha$, is formulated as
$$\alpha =\min [\alpha_0 (h/r)^n, \alpha_{\max}],\eqno(7)$$
where $\alpha_0$ is a constant and $\alpha_{\max} (\leq 1)$ represents the
effect of saturation of the anomalous turbulence.

In the radiative diffusion and vertically average approximation,
the viscosity, $\bar{\nu}$, is calculated by using the
half-thickness of the disk, $H$, as the scale height $h$ and
$\sqrt{2Hp/\Sigma}$ as the sound speed $c_s$.
This form of viscosity leads to a viscous stress corresponding to the
$\tau=-\alpha p_g$ model in the Keplerian approximation, and it has a
corresponding heating rate of
$$\bar{F}^+ = \bar{\nu}\Sigma {\left(r{\pd \Omega r}\right)}^2.\eqno(8)$$
In this approximation, the cooling rate can be expressed as
$$\bar{F}^- = {4acT^4 \o 3 \kappa \Sigma} \eqno(9)$$
where $\kappa$ is the sum of electron scattering and free-free opacities.

The heating and cooling rates in the energy equations can also be calculated
directly from the detailed vertical structures of the disk. In practice
in the time dependent study
of dwarf nova accretion disk models, the above simplified heating and cooling
rates in equations (8) and (9) are usually not applied directly because they
differ
from the vertical structure results qualitatively due to the temperature
sensitivity of the opacity.
Many schemes have been introduced to deal with this
situation (see, \eg, Papaloizou, Faulkner \& Lin 1983; Mineshige \& Osaki 1983;
Meyer \& Meyer-Hofmeister 1984; Cannizzo, Wheeler \& Polidan 1986; Mineshige
1986). For example, Faulkner, Lin \& Papaloizou (1983)
and Papaloizou, Faulkner \& Lin (1983) used a modified approximate expression
for the cooling rate. Mineshige \& Osaki
(1983) tabulated both the cooling and heating rates from detailed vertical
structures. To construct the disk vertical
structures, four
equations (which are, in the steady state approximation,
hydrostatic equilibrium, energy transport and energy conservation equations)
are required to obtain the solutions of density,
temperature and vertical energy flux with respect to the height from the disk
mid-plane. In the approach of Mineshige and Osaki (1983),
the condition of thermal equilibrium in the
vertical direction (\ie, the energy conservation equation) is relaxed, and
instead, a relationship between the energy flux and the height is assumed.

For disks surrounding black holes and neutron stars, MCT constructed detailed
steady state vertical structures including convection and showed that
convection has a stabilizing effect on the disk and changes the disk structure
significantly. Therefore, we choose to use the cooling and heating rates from
the detailed vertical structure models instead of the simplified rates above.
To tabulate the cooling and heating rates, a different approach from that of
Mineshige \& Osaki (1983) is adopted as follows.

We assume that the departure from thermal equilibrium in the vertical
direction has a form of
$$ F^- = \delta F^+,\eqno(10)$$
where, $\delta$ is a dimensionless constant and is introduced to represent the
effect of advection of energy by material motion and radiative energy
transport in the radial
direction. For $\delta >1$, the radial energy transport is a heating process,
and for $\delta <1$, it is a cooling process. The energy equation in the
vertical direction becomes
$$\deriv F z = \delta \rho \epsilon_{\nu}, \eqno(11)$$
where $z$ is the height from the disk mid-plane, $F$, $\rho$ and
$\epsilon_\nu$  are the total flux, the density and the viscous energy
generation rate at height $z$ respectively. Note that, in the calculation of
$\epsilon_{\nu}$, which is $9/4 \nu \Omega^2$, the viscosity is computed with
the local adiabatic sound speed and a modified local scale
height and they are functions of $z$ (see MCT).
Now, if we identify the temperature $T(z)$ at
the mid-plane as $T_c$, the total flux $F(z)$ at the surface as $F_s$ and
the column density $\Sigma(z)$ at the surface as $\Sigma_s$, then we have the
following relations for $\Sigma$, $T$, $F^-$ and $F^+$:
$$\left\{ \begin {tabular}{l}
   $\Sigma=2\Sigma_s,$\\[0.5ex]
   $T=T_c,$\\[0.5ex]
   $F^-=2F_s,$\\[0.5ex]
   $F^+=2F_s/\delta.$
   \end{tabular} \right. \eqno(12)$$
In general, $\delta$ is unknown and $F^-$ and $F^+$ can not be obtained
from $\Sigma$ and $T$. However, for a given viscosity prescription and
specified radius, it is found that two relations for $F^-$ and $F^+$
exist which are approximately independent of $\delta$.  These relations can be
formally expressed as
$$\left\{ \begin {tabular}{l}
   $f_1(\ln{\bar{F}^-}) = \ln {F^-}, $\\[0.5ex]
   $f_2(\ln{\bar{F}^-}) =  F^+/\bar{F}^+.$\\
   \end{tabular} \right. \eqno(13)$$
As an example, $f_1$ and $f_2$ are shown in Figure 1(a) and (b) respectively
for a specified set of model parameters and radius. In each plot, there are
three curves of $\delta=0.8,1.0$ and 1.5 (corresponding
to dotted, solid and dashed lines respectively), however,
the difference between them is very small. The $f_2$ curve is Z-shaped, and is
related to the S-shaped relation between the temperature and surface
density of the disk which will be discussed in the following sections.

These two functions are tabulated with respect to $\bar{F}^-$ at specified
radii for a given viscosity prescription.  For the time dependent evolution,
$\bar{F}^-$ and $\bar{F}^+$ are calculated from the two variables
$\Sigma$ and $T$ (see eqs. [8] and [9]), and linear interpolation is used to
obtain $f_1$ and $f_2$ from the tables to determine the modified values of
$F^-$ and $F^+$, \ie,
$$\left\{ \begin {tabular}{l}
   $F^- = \exp(f_1),$\\[0.5ex]
   $F^+ = f_2 \bar{F}^+.$\\
   \end{tabular} \right. \eqno(14)$$

The time dependent equations (1) and (2) are solved to calculate $\Sigma$ and
$T$ via an explicit method. The accretion disk is divided into 41 grid points
distributed equally on logarithmic scale ranging from an inner boundary at
$4\,r_g$ to an outer boundary at $300\,r_g$, where $r_g$ is the Schwarzschild
radius ($r_g = 2 G M/c^2$). The initial structure of the disk is given by
the thermal equilibrium solutions ($\delta=1$)
of the detailed vertical structures (see MCT). At every grid radius
a series of models of steady state
vertical structures with different mass accretion rates are constructed, and
at the same time, functions of $f_1$ and $f_2$ are tabulated. For the initial
values of $\Sigma$ and $T$ with the same steady state mass accretion rate at
each grid point, linear interpolation
from the vertical structure models of the same grid radius is applied. The
boundary conditions are chosen to correspond to zero gradients at the innermost
radius and to a fixed temperature and surface density at the outermost radius.

The disk model is determined by the mass of the central object, $M$, the mass
accretion rate, $\mdot$, and the viscosity parameters, $\alpha_0$, $n$, and
$\alpha_{\max}$. We adopt the mass as
$1.4\,\msun$ for neutron stars and $10\,\msun$ for black holes.
The mass accretion rate is measured in units of the Eddington limit defined as
$\mdote = 4 \pi G M/ (\kappa_e c \epsilon)$,
where $\kappa_e$ is the electron scattering opacity and
$\epsilon$ is the efficiency for the conversion of rest mass energy
into radiation taken to be 1/6 for a neutron star and 1/16 for a black hole.

\bigskip

\cl{\bf 3. THERMAL AND VISCOUS INSTABILITIES}
\medskip
\par
Thermal instability occurs when the heating process is more efficient than the
cooling process. In an optically thick and geometrical thin viscous accretion
disk, if we consider only the local processes, then the heating is due to the
viscous dissipation and the cooling is due to the radiative diffusion (the
convection is neglected). The general criterion for thermal instability may
be expressed as (see Piran 1978)
$$\left.{\pd {\ln \bar{F}^+} {\ln H }}\right|_{\Sigma}
 >\left.{\pd {\ln \bar{F}^-} {\ln H}} \right|_{\Sigma}.\eqno(15)$$

The local thermal instability of accretion disks for which the
viscous stress is given in the form of $\tau = -\alpha_0 (h/r)^n p_g$ has
been examined by Meyer (1986) in the limit in which radiation pressure
dominates gas pressure and it was shown that thermal
instability occurs for $n > 0.75$.
In a recent study, MCT generalized the analysis to arbitrary values
of the ratio of gas pressure to total pressure ($\beta$) and derived the
following condition for thermal instability:
$$ \beta < {4n-3 \o 3(1+n)}=\beta_c.\eqno(16)$$
Furthermore, it was shown that this condition
is identical to the viscous instability condition which can be inferred
from the slope of the relation between the mass accretion rate and surface
density at a fixed radius. That is, for $\deriv \mdot \Sigma > 0$ the disk is
locally stable whereas for $\deriv \mdot \Sigma < 0$ it is locally unstable
(see Bath \& Pringle 1982).
In particular, in the polytropic approximation, the slope is given as (see
MCT)
$$\deriv {\ln \mdot} {\ln \Sigma} = - {{5 + (5+n)\beta}\over {4n-3-3(n+1)
\beta}}. \eqno(17)$$
{}From equation (17), for a given $n$ (less than 6), it is seen that, in the
parameter plane of mass accretion
rate and surface density, the slope of $\mdot(\Sigma)$ relation
is positive if the gas pressure is
much larger than the radiation pressure, and which corresponds to the stable
regime. At the point $\beta=\beta_c$, the slope changes
sign from ${\deriv\mdot \Sigma} > 0$ to ${\deriv\mdot \Sigma} < 0$, and the
condition of $\beta < \beta_c$ corresponds to the unstable regime.

In the detailed vertical structures of accretion disks, the above feature is
qualitatively recovered. However, another turning point was revealed, which is
due to the saturation of the $\alpha$ parameter (see MCT), and, hence, an
S-shaped curve is formed in the plane of mass accretion rate and surface
density. The upper branch is stable because in that regime
($\deriv {\ln \mdot} {\ln \Sigma} \simeq 5/3$) the disk is
approximately described by the $\tau = -\alpha_{\max} p_g$ prescription, and no
dependence of the viscous parameter on the local scale height exists.

An S-shaped relation of temperature with respect to surface density follows
from the relation between $T$ and $\mdot$, specifically,
$$\mdot \propto F^- \propto T^4/\Sigma. \eqno(18)$$
It is also seen (eq. [17]) that, for larger $n$, $|{\deriv \mdot \Sigma}|$ or
$|{\deriv T \Sigma}|$ at the middle branch of the S-curve is smaller and
so the S-curve becomes steeper, which results in a greater tendency toward
instability. This
can also be explained through the different sensitivities of the cooling and
heating rates on
the temperature. For the the viscous heating rate, we have, approximately,
$$\left.{\pd {\ln \bar{F}^+} {\ln T }}\right|_{\Sigma}
\simeq (2+n){4-3\beta \o 1+\beta}-{7-7\beta \o 1+\beta}, \eqno(19)$$
which becomes larger for larger $n$. On the other hand,
$$\left.{\pd {\ln \bar{F}^-} {\ln T }}\right|_{\Sigma}
\simeq 4, \eqno(20)$$
so the temperature sensitivity of the cooling rate is approximately a constant.

A local thermal-viscous instability at a fixed radius follows from the
evolutionary trajectory in the $(T,\Sigma)$ plane. For example, as the mass
accretion rate increases while the disk is on the lower stable branch,
the surface
density increases until the turning point where $\beta=\beta_c$ is reached.
The subsequent evolution is expected to lead to heating of the disk until
the upper stable branch is reached. The mass flow rate corresponding to this
state, however, is higher than that corresponding to the lower stable branch
and therefore $\Sigma$ decreases. Eventually, the evolutionary path reaches
the upper turning point and the disk undergoes a transition to the lower
stable branch, where the mass flow rate is less than the assumed steady state
rate. As a result, matter accumulates, thereby increasing $\Sigma$ and the
cycle begins anew. To determine whether the disk follows the steady state
disk curves, as outlined above, we turn to the description of the global
evolution.

\bigskip

\cl{\bf 4. GLOBAL EVOLUTION}
\medskip
\par
The parameters characterizing each of the model sequences as well
as the results of the simulations are summarized in Table~1. Here,
$\mdot$ is in units of the Eddington value. For the unstable models,
the frequency
of the oscillation, $f$, and the relative amplitude of the luminosity
fluctuation, $\Delta L/L$, are also listed.

\bigskip

\cl{\bf 4.1 Neutron Stars}
\medskip
\par
The existence of an S-shaped relation between $T$ and $\Sigma$ at a
specified radius indicates thermal and viscous instabilities whenever the
steady state solutions of $T$ and $\Sigma$ are located at the
middle branch. However, these instabilities are local and may be stabilized
by nonlinear effects especially if the instability is restricted
to only a small region of the disk. Here,  we present the results of the
time dependent
calculation of model sequence~1 ($n=1.1$) with a
mass accretion rate of $ 0.3\,\mdote$ and a central object of mass equal to
$1.4\,\msun$ to demonstrate the global stabilizing
effects. The steady state (\ie, the initial state) is indicated by open
circles in the $T$ and $\Sigma$ plane at four specified radii
(see Fig.~2). The ratios of gas to total pressure, $\beta$, are also listed
for each radius.
In the radiative diffusion approximation, for $n=1.1$, the critical value
of the ratio is, $\beta_c=0.2121$ (see eq. [16]). Thus,
the disk is predicted to be unstable in three of the four radii displayed.
However, it is seen that, only two open circles in the inner
radii are on the unstable middle branches, which confirms the results of MCT
that the convection has stabilizing effects and a greater contribution of
radiation pressure is
required to make the disk locally unstable. The nonlinear calculation
shows further that the global effects, \ie, the advection of energy
and radiation
transport in radial direction, stabilize the entire disk, even though it is
locally unstable in the inner regions. As is shown
in Figure~2, the evolution models are seen to be
approximately the same as that of the initial steady state (the
last evolution model is indicated by the cross). The variation of disk
luminosity, calculated by integrating the flux $F^-$ over the area of the
disk surface, with respect to time is displayed on Figure~3. Obviously, it
evolves to a steady
state after a short initial transient stage of less than 10 seconds.
Additional calculations have been performed for different mass accretion
rates ranging from 0.01 to 1 times the Eddington value, and the disk is always
found to evolve to a stable steady state structure.
These results do not confirm local analysis which predicts
instability for $n \gapprox 0.75$ and reveal that stabilization for
$n \lapprox 1.1$ results from the global effects of non-local energy transport
(see also Taam \& Lin 1984).

The behavior of the accretion disk is a sensitive function of the
parameters characterizing the viscosity prescription (see \S3).
To strengthen the disk instability, we increase the viscosity
parameter index $n$ to 1.3 and keep the other viscosity parameters fixed as
that of
model sequence~1. We denote that model as sequence~2. The results for
$\mdot=0.27\,\mdote$
are shown in Figure~4
and Figure~5a. It is seen that, the evolution paths of $T$ and $\Sigma$ at
radii
$5.53\,r_g$ and $11.77\,r_g$ ( see Figs.~4a,\,b) are both very narrow loops
moving in a counterclockwise direction.
The narrow loops represent a weak instability, which can also be seen
from the small amplitude periodic oscillations of the disk luminosity
(Fig.~5a).
In particular, the frequency of the oscillation is about 0.143\,Hz and
the amplitude is about $22\%$. At the other two larger radii,
$31.10\,r_g$ and $59.43\,r_g$, the steady state solutions of $T$ and $\Sigma$
are located at the lower branch of the S-curve (see Fig.~4c,\,d), and they
are stable. The instability becomes stronger at higher
mass accretion rates ($\mdot = 0.3 \mdote$)
because radiation pressure becomes more important and
the unstable regions of the disk is wider. Therefore, a
lower frequency, larger amplitude oscillation
results (see Fig.~5b). On the other hand, for mass accretion rates less than
$0.21\,\mdote$, the disk is eventually stablized (Fig.~5c).

The low frequency QPOs of about 0.04\,Hz from the Rapid Burster
imply a relatively small effective $\alpha$ parameter.
To obtain a smaller effective $\alpha$ parameter, one can either increase $n$
or
decrease $\alpha_0$. However, for a decrease in $\alpha_0$,
the surface density increases and, for a given $\mdot$, radiation pressure
becomes less important in the disk. Instability can only, therefore, occur at
higher mass accretion rates. Since the persistent mass accretion
rate in the Rapid Burster is probably low ($\lapprox 0.2\,\mdote$), a
larger $n$ is suggested. Furthermore,
the parameter $\alpha_{\max}$ determines the range of mass accretion rates over
which the disk is unstable.  For smaller $\alpha_{\max}$, the range of mass
accretion rates is smaller. The absence of low frequency QPOs during the
type II X-ray bursts may require a smaller $\alpha_{\max}$.

In model sequences~$5 - 8$, $n=1.6$, $\alpha_0=30$ and $\alpha_{\max}=0.2$ are
applied.
The larger value of $n$ makes the disk more locally unstable and also results
in a wider unstable region of the disk. This effect, in addition to a smaller
effective viscosity, leads to a longer viscous time scale.
We first examine sequence~5 with $\mdot= 0.14 \mdote$.
Figure~6 shows the evolution path in
the temperature and surface density space at four specified radii. As
expected, at the inner radii, the path has larger loops because of a stronger
instability. However, the evolution path does not jump vertically
from the first
turning point to the upper branch. Instead, the trajectory takes an
intermediate route
between the upper and middle branches (closer to the latter one).
The evolution path returns to the lower
branch by following a route also closer to the middle branch. On the other
hand,
it does follow the lower branch approximately because the effects of radial
advection are negligible there.
At a larger radius, the loop is smaller, and eventually, it dissapears.
It should be noticed that, in this case, the unstable region is wider than
indicated by the local instability analysis. As seen in Figure~6c, at radius
$31.10\,r_g$, the initial
model is located at the lower stable branch, however, due to the global
radial motions, perturbations in the inner regions move outwards beyond this
point and the stable region becomes unstable. In this model sequence,
the unstable region of the disk is spatially confined inside $40\,r_g$.

The light curve resulting from the mass flow modulations in the
accretion disk is illustrated in Figure~7a.  The period
of the oscillations is found to be $\sim$ 21.5\,s and the amplitude of the
fluctuations correspond to ${\Delta L \o L} \sim 67\%$.

To determine the sensitivity of the results to the mass accretion rate,
the mass accretion rate was decreased to $\mdot/\mdote=0.12$ in
sequence~6. The disk is unstable as evidenced by the luminosity
variations illustrated in Figure~7b. The fluctuations are manifested as
oscillations and are similar to Figure~7a. It
is found that the period has decreased to 11\,s, and
the relative luminosity amplitude has decreased,
to $\sim 20\%$. However, the relation that the frequency
of the oscillation decreases and the relative luminosity fluctuation
amplitude increases as the mass accretion rate is increased, holds only
over a limited range in mass accretion rates for which the disk is unstable.
For higher mass accretion rates, the amplitude of the oscillations
may decrease and eventually the disk becomes stable due to the saturation of
the viscosity parameter, $\alpha$. For a set of viscosity parameters
$(n,\alpha_0,\alpha_{\max})$ $\sim$ $(1.6,30,0.2)$,
it is found that the accretion disk is stable for $\mdot \lapprox 0.11 \mdote$
(sequence~8) and for $\mdot \gapprox 0.47\mdote$ (sequence~9).

For a large viscosity parameter index, such as $n=1.6$, the variation of the
disk luminosity may exhibit different evolution patterns in some range of
mass accretion rates. One example is shown in Figure~7c where two local
maxima can be seen.

\bigskip

\cl{\bf 4.2 Black Hole Candidates}
\medskip
\par
To determine the dependence of the results on the mass of the compact object
we increased $M$ to 10 $\msun$ to model the evolution of an accretion disk
surrounding a black hole.  In order to produce oscillations at frequencies
$\sim 0.04$ Hz, the effective viscosity parameter, $\alpha$, must be larger
than that
of the neutron star case since the absolute size of the inner disk is larger.
We first keep $\alpha_0=30$ unchanged, but use $n=1.2$
(see model sequences~10--13, Table~1).
The instability is moderately mild, as shown by the disk luminosity
oscillations in Figure~8a. For example, in the case of sequence~10 with mass
accretion rate of $0.076\,\mdote$, the
oscillation has a frequency of 0.026\,Hz, and a small amplitude of
$\Delta L /L \sim 0.083$. For an increase of mass accretion rate, the
frequency of the oscillation becomes even lower (sequence~11). The range
of mass accretion rates in which the disk is unstable is about a factor of 3,
specifically, $0.07 \lapprox \mdot/\mdote \lapprox 0.20$ (sequences~12 and 13).
In order to obtain a higher frequency, \ie, 0.04\,Hz, $\alpha_0=50$ is applied
in sequence~14--16. The frequency of the luminosity oscillations of
sequence~14 with mass accretion rates $0.076\,\mdote$ is increased to
0.035\,Hz (see Fig.~8b). However, the range of mass accretion rates for
which instability is
indicated is very narrow in comparision with the model sequences with
$\alpha_0=30$ (\ie, sequences~12--13), specifically,
$0.071 \lapprox \mdot/\mdote \lapprox 0.084$.
The smaller upper limit on the
mass accretion rate is due to the larger $\alpha_0$ which makes the viscosity
parameter saturate more easily.

The mild instability indicated by $n=1.2$
suggested that this value of the viscosity parameter index is near the lower
limit for an unstable disk.
This is confirmed by model sequence~17, which is performed to shown that
$n=1.1$ is stable for a $10\msun$ black hole in a wide range of mass accretion
rates, specifically,
$0.005 \lapprox \mdot/\mdote \lapprox 2.0$.
For a larger $n$, the strength of the instability become
stronger and for a fixed $\alpha_0$ the period of the oscillation is longer.
For a $10\,\msun$ black hole, the possible modulation frequency of
oscillations of the disk luminosity
may always be lower than the 0.04\,Hz QPOs observed. Thus, we repeated the
simulations of
the models with viscosity parameters of $n=1.2$, $\alpha_0=30$, and
$\alpha_{\max}=1.0$, but with a $5\,\msun$ black hole. In this case, the range
of mass accretion rates in which the disk is unstable is narrower than in
that of a $10\,\msun$ black hole (see sequences~12--13), specifically,
$0.085 \lapprox \mdot/\mdote \lapprox 0.18$.
The frequency of the oscillations, as expected, is higher.
An example is given in model sequence~18 with $\mdot=0.1\mdote$, where,
the frequency is about 0.05\,Hz.

In the case of a $5\,\msun$ black hole, a wider range
of mass accretion rates in which the disk is unstable can be obtained
with $n=1.3$ (sequence~19--21), and the time scale of the oscillation is
in the required range, \ie, $\sim 0.04$\,Hz (see Fig.~8c for sequence~19)

\bigskip

\cl{\bf 5. DISCUSSION}
\medskip
\par
We have demonstrated by global analysis that accretion in a geometrically
thin, optically thick disk surrounding either a neutron star or a black hole
can be unstable.  The instability is thermal in origin and results from the
inability of the disk to maintain a local thermal balance whenever radiation
significantly contributes to the pressure in the disk. The strength of the
instability is determined by the sensitivity of the viscous
heating rate to the temperature, and is greater for larger $n$.
The instability may be restricted to a relatively narrow spatial extent in
the disk, and the nonlinear calculations reveal that the instability results
in luminosity oscillations rather than bursts.
Such oscillations may exhibit a quasi-periodic behavior if the mass flow rate
entering the inner region of the disk is not strictly constant.

The thermal-viscous instability of the accretion disks can be understood
through the steady state $T \sim \Sigma$ relation curve (or effectively,
the relation between $\mdot$ and $\Sigma$). The S-shaped relation
for $T(\Sigma)$ exists in slim accretion disk models (see \eg, Abramowicz,
Czerny, Lasota \& Szuszkiewicz 1988; Szuszkiewicz 1990; Chen \& Taam 1993),
with its upper stable branch due to radial advection, which is a
cooling process. The stabilization occurs only at super-Eddington rates.
The S-shaped relation of $T$ and $\Sigma$ is also well known in disk
models of dwarf novae. In that case, the stable upper branch represents a
nearly fully ionized radiative disk structure with the gas pressure dominant.
The middle unstable branch is due to the
partial ionization of the material, which results in a sensitive temperature
dependent opacity. The lower stable branch corresponds to cool non-ionized
disk structure. It has been commonly assumed that the evolutionary
path in $T$ and
$\Sigma$ follows the stable lower and upper branches but does not follow
the unstable middle branch. Instead, the path makes a vertical transition at
constant $\Sigma$ from the lower turning point
to the upper branch, and a downward transition from the upper
turning point to the lower branch. This assumption is usually confirmed by
the global time dependent calculations (see review by Cannizzo, 1993).
However, it is not obvious that is always the case because the evolution path
may
not neccessarily follow the steady state curve, which is a result of the
vertical thermal equilibrium. For example, the evolution path will
be affected by the radial motion of the disk which could even stablize it.
This has been demonstrated by our results and, in fact, it has been also shown
by the calculations of Honma \etal (1991) in the slim disk approximation.
Our results show that, disks with viscosity parameter index $n \lapprox 1.1$
are globally stable. Our
results also reveal that the evolutionary paths in the $T$ and $\Sigma$ plane
are loops and the size of the loop reflects the strength of
the global instability of the disk.

Even though the time scales of the thermal and viscous instability
at different radii of the disk are different,
the global evolutions of the disk show that the evolution time scale is not
determined locally and the instabilities are globally coherent. To see this,
we plot the time variations of the radial distribution of the mass accretion
rate of model sequence~19 (see Table 1) in Figure~9.
The instability is initiated in the inner region where the
contribution of radiation to the total pressure is the greatest and, as a
result,
a large amount of mass moves inwards at a high radial velocity.
This results in a low surface density and a transition wave which
propagates outward. This wave terminates at the point where the disk is
stable.
In the next stage of the evolution the mass in the stable region of the
disk diffuses inwards, and the surface density of the inner region begin to
increase, as does the temperature. Eventually, the instability is
triggered anew.

The unstable behavior is found to lie in a range of mass accretion rates,
$\mdot_{c1} \lapprox \mdot$ $\lapprox \mdot_{c2}$,  where $\mdot_{c1}$ and
$\mdot_{c2}$ depends on $\alpha_0$ and $\alpha_{\max}$ for a given $n$.
For larger $\alpha_0$
and/or smaller $\alpha_{\max}$, the difference between $\mdot_{c2}$ and
$\mdot_{c1}$ decrease.  For example, for the model parameters adopted (see
Table 1), $\mdot_{c1}$ and $\mdot_{c2}$ would be 0.11 and 0.47 $\mdote$ for
a neutron star and 0.045 and 0.38 $\mdote$ for a black hole of $5\,\msun$.

Within the framework of the disk instability model, it is a general property
of these instabilities that the frequency of the oscillation is a function
of the mass accretion rate, with the frequency increasing for lower source
intensity (assuming a direct relation between intensity and mass accretion
rate).  Such a behavior appears to have been observed in the Rapid Burster
(MXB~1730--335) by Lubin et al. (1992) in the hump immediately
following the post dip phase of some Type II bursts.  In addition, Lubin et
al. (1992) find that the amplitude of the oscillations ($\lapprox 60\%$)
decreased as the persistent flux declined.  This is also consistent with
the general trends exhibited by the numerical results in the previous section.
Furthermore, the observations suggest that the level of persistent emission
in the Rapid Burster corresponds very closely to the minimum accretion rate
necessary for instability, $\mdot_{c1}$, since the oscillations disappear as
the persistent emission decreases. If we identify the persistent luminosity
level to correspond to $\mdot \sim 0.14 \mdote$, the low frequency (0.04 Hz)
and large amplitude of luminosity oscillations imply a large $n$ ($\sim 1.6$)
and constrain $\alpha_0$ to be $\sim 30$.
This value for $n$ is similar to that inferred for a viscosity based on
magnetic dynamo involving non-uniform rotation and
cyclonic convection (Meyer \& Meyer-Hofmeister 1983).
The limit on $\alpha$ (i.e.
$\alpha_{\max}$) may be constrained by the absence of these oscillations
during the bump phase observed on the decline from a Type II burst (see Lubin
et al. 1993).  The intensity ratio of the bump phase to the persistent phase
limits the range of mass accretion rates to $\lapprox 2.5$.  This suggests
that $\alpha_{\max} \lapprox 0.2$.

As shown in \S4 the properties of the calculated disk oscillations are
sensitive to the form of the viscosity parameters.  Hence, the observations
of oscillations from other galactic X-ray binary sources can
provide additional constraints on the
unknown parameters.  Low frequency oscillations $\sim 0.04$ Hz have also been
observed from the black hole candidate sources Cyg X-1 (Angelini \& White
1992; Vikhlinin \etal 1992a, 1994; Kouveliotou \etal 1992a) and GRO~J0422+32
(Vikhlinin \etal 1992b; Kouveliotou \etal 1992b; Pietsch \etal 1993), which
suggests that $n \sim 1.2-1.3$ and $\alpha_0 \sim 30$ for a $5 \msun$ black
hole. These lower values for $n$ are close to those expected for a viscosity
based on internal wave driven dynamo (Vishniac \& Diamond 1992).
Further
long term observations, such as those carried out for the Rapid Burster,
will be required to provide additional constraints on these parameters as
well as on $\alpha_{\max}$.

Finally, we remark that within the framework of the model proposed here
one would expect that other X-ray binaries would also exhibit low frequency
QPO phenomenon provided that the presence of a magnetosphere about a neutron
star does not exclude the presence of the unstable region in the Keplerian
disk.  In those systems where the accretion disk extends to the compact
object the instability may be restricted to a small range of mass accretion
rates (as in the Rapid Burster), thus making it difficult to determine
whether a given source can exhibit such phenomenon.  In this case, transient
sources would be ideal candidates for the search of such QPOs since
the level of intensity varies over a wide range.

\vskip 0.2 in

The authors are grateful to A. Vikhlinin for sending a copy of his
paper before publication.
This research has been supported in part by NASA under grant NAGW-2526.

\newpage
\begin {tabular}{lccccccc}
\multicolumn{8}{c}{{\bf Table 1}}\\[1.5ex]
\multicolumn{8}{c}{Model Sequences}\\[1.0ex] \hline\hline \\[-1ex]
Sequence & $M(\msun)$ & $\mdot$ & $n$  & $\alpha_0$ & $\alpha_{\max}$
& $f$ (Hz)  & ${\Delta L \o L}$ \\[1ex] \hline\\[-1.5ex]
1............ &  1.4 & 0.30   & 1.1 & 20 & 1.0 & ....  & 0.0 \\[0.5ex]
2............ &  1.4 & 0.27   & 1.3 & 20 & 1.0 & 0.143 & 0.22\\[0.5ex]
3............ &  1.4 & 0.30   & 1.3 & 20 & 1.0 & 0.125 & 0.31\\[0.5ex]
4............ &  1.4 & 0.21   & 1.3 & 20 & 1.0 & ....  & 0.0 \\[0.5ex]
5............ &  1.4 & 0.14   & 1.6 & 30 & 0.2 & 0.046  & 0.67\\[0.5ex]
6............ &  1.4 & 0.12   & 1.6 & 30 & 0.2 & 0.093  & 0.20\\[0.5ex]
7............ &  1.4 & 0.125  & 1.6 & 30 & 0.2 & 0.037  & 0.27\\[0.5ex]
              &      &        &     &    &     & 0.037  & 0.12\\[0.5ex]
8............ &  1.4 & 0.11   & 1.6 & 30 & 0.2 & ....  & 0.0 \\[0.5ex]
9............ &  1.4 & 0.47   & 1.6 & 30 & 0.2 & ....  & 0.0 \\[0.5ex]
10......... & 10.0 & 0.076  & 1.2 & 30 & 1.0 & 0.026 & 0.083 \\[0.5ex]
11......... & 10.0 & 0.10   & 1.2 & 30 & 1.0 & 0.019 & 0.31 \\[0.5ex]
12......... & 10.0 & 0.07   & 1.2 & 30 & 1.0 & ....  & 0.0 \\[0.5ex]
13......... & 10.0 & 0.20   & 1.2 & 30 & 1.0 & ....  & 0.0 \\[0.5ex]
14......... & 10.0 & 0.076  & 1.2 & 50 & 1.0 & 0.035 & 0.056 \\[0.5ex]
15......... & 10.0 & 0.085  & 1.2 & 50 & 1.0 & ....  & 0.0 \\[0.5ex]
16......... & 10.0 & 0.071  & 1.2 & 50 & 1.0 & ....  & 0.0 \\[0.5ex]
17......... & 10.0 & 0.005--2.0 & 1.1 & 20 & 1.0 & ....  & 0.0 \\[0.5ex]
18......... & 5.0 & 0.10    & 1.2 & 30 & 1.0 & 0.05  & 0.12 \\[0.5ex]
19......... & 5.0 & 0.06    & 1.3 & 30 & 1.0 & 0.036 & 0.35 \\[0.5ex]
20......... & 5.0 & 0.045   & 1.3 & 30 & 1.0 & ....  & 0.0 \\[0.5ex]
21......... & 5.0 & 0.38    & 1.3 & 30 & 1.0 & ....  & 0.0 \\ \hline
\end {tabular}

\newpage
\cl{\bf REFERENCES}
\medskip
\ref Abramowicz, M. A., Czerny, B., Lasota, J. P., \& Szuszkiewicz, E. 1988,
\apj{332}, 646
\smallskip

\ref Abramowicz, M. A., Szuszkiewicz, E., \& Wallinder, F. 1989, in Theory
of Accretion Disks, ed. F. Meyer, W. J. Duschl, J. Frank, \&
E. Meyer-Hofmeister (Dordrecht: Kluwer), 141
\smallskip

\ref Alpar, M. A., Hasinger, G., Shaham, J., \& Yancopoulos, S. 1992,
\aaa{257}, 627
\smallskip

\ref Alpar, M. A., \& Shaham, J. 1985, \nature{316}, 239
\smallskip

\ref Angelini, L., \& White, N. 1992, IAU Circ. 5580
\smallskip

\ref Bath, G. T., \& Pringle, J. E. 1982, \mnras{199}, 267
\smallskip

\ref Cannizzo, J. K. 1993, in Accretion Disks in Compact Stellar Systems,
ed. J. Craig Wheeler (Singapore: World Scientific Publishing), in press
\smallskip

\ref Cannizzo, J. K., Wheeler, J. C. \& Polidan, R. S. 1986, \apj{301}, 634
\smallskip

\ref Chen, X., \& Taam, R. E. 1993, \apj{412}, 254
\smallskip

\ref Coroniti, F. V. 1981, \apj{244}, 587
\smallskip

\ref Duschl, W. J., \& Livio, M. 1989, \aaa{209}, 183
\smallskip

\ref Faulkner, J., Lin, D. N. C., \& Papaloizou 1983, \mnras{205}, 305
\smallskip

\ref Fortner, B., Lamb, F. K., \& Miller, G. S. 1989, \nature{342}, 775
\smallskip

\ref Honma, F., Matsumoto, R., \& Kato, S. 1991, \pasj{43}, 147
\smallskip

\ref Kouveliotou, C., \etal 1992a, IAU Circ. 5576
\smallskip

\ref Kouveliotou, C., \etal 1992b, IAU Circ. 5592
\smallskip

\ref Lamb, F. K., Shibazaki, N., Alpar, M. A., \& Shaham, J. 1985,
\nature{317}, 681
\smallskip

\ref Lasota, J. P. \& Pelat, D. 1991, \aaa{249}, 574
\smallskip

\ref Lewin, W. H. G., van Paradijs, J., \& van der Klis, M. 1988, \ssr{46}, 273
\smallskip

\ref Lightman, A. P., \& Eardley, D. N. 1974, \apj{187}, L1
\smallskip

\ref Lubin, L. M., Lewin, W. H. G., Rutledge, R. E., van Paradijs, J.,
van der Klis, M., \& Stella, L. 1992, \mnras{258}, 759
\smallskip

\ref Lubin, L. M., Lewin, W. H. G., van Paradijs, J., \& van der Klis
1993, \mnras{261}, 149
\smallskip

\ref Lynden-Bell, D., \& Pringle, J. E. 1974, \mnras{168}, 603
\smallskip

\ref Matsumoto, R., Kato, S., \& Honma, F. 1989, in Theory of Accretion
Disks, ed. F. Meyer, W. J. Duschl, J. Frank, \& E. Meyer-Hofmeister
(Dordrecht: Kluwer), 167
\smallskip

\ref Meyer, F. 1986, in Radiation Hydrodynamics in Stars and Compact Objects,
ed. D. Mihalas \& K. -H. A. Winkler (Berlin: Springer-Verlag), 249
\smallskip

\ref Meyer, F., \& Meyer-Hofmeister, E. 1983, \aaa{128}, 420
\smallskip

\ref Meyer, F., \& Meyer-Hofmeister, E. 1984, \aaa{132}, 143
\smallskip

\ref Milsom, J. A., Chen, X., \& Taam, R. E. 1994, \apj{421}, in press (MCT)
\smallskip

\ref Mineshige, S. 1986, \pasj{38}, 831
\smallskip

\ref Mineshige, S., \& Osaki, Y. 1983, \pasj{35}, 377
\smallskip

\ref Nowak, M. A., \& Wagoner, R. V. 1991, \apj{378}, 656
\smallskip

\ref Papaloizou, J. C. B., Faulkner, J., \& Lin D. N. C. 1983, \mnras{205}, 487
\smallskip

\ref Papaloizou, J. C. B., \& Pringle, J. E. 1984, \mnras{208}, 721
\smallskip

\ref Papaloizou, J. C. B., \& Pringle, J. E. 1985, \mnras{213}, 799
\smallskip

\ref Papaloizou, J. C. B., \& Pringle, J. E. 1987, \mnras{225}, 267
\smallskip

\ref Pietsch, W., Haberl, F., Gehrels, N., \& Petre, R. 1993, \aaa{273}, L11
\smallskip

\ref Piran, T. 1978, \apj{221}, 652
\smallskip

\ref Shakura, N. I., \& Sunyaev, R. A. 1973, \aaa{24}, 337
\smallskip

\ref Shakura, N. I., \& Sunyaev, R. A. 1976, \mnras{175}, 613
\smallskip

\ref Spruit, H. C. 1987, \aaa{184}, 173
\smallskip

\ref Stella, L., \& Rosner, R. 1984, \apj{277}, 312
\smallskip

\ref Szuszkiewicz, E. 1990, \mnras{244}, 377
\smallskip

\ref Taam, R. E., \& Lin, D. N. C. 1984, \apj{287}, 761
\smallskip

\ref van der Klis, M. 1989, \annrev{27}, 517
\smallskip

\ref Vikhlinin, A., \etal 1992a, IAU Circ. 5576
\smallskip

\ref Vikhlinin, A., \etal  1992b, IAU Circ. 5608
\smallskip

\ref Vikhlinin, A., \etal  1994, ApJ, in press
\smallskip

\ref Vishniac, E. T., \& Diamond, P. H. 1992, \apj{398}, 561
\smallskip

\ref Wallinder, F. H. 1991, \aaa{249}, 107
\smallskip

\newpage
\cl{\bf FIGURE CAPTIONS}
\bigskip

\n{\bf Figure 1.} Curves of $f_1$ (a) and $f_2$ (b) used to tabulate
the cooling and heating rates. The dotted, solid and dashed lines,
correspond to the thermal equilibrium departure parameter,
$\delta$, of 0.8, 1.0, 1.5 respectively. In this example, the central object
is a $10 \msun$ black hole, the viscosity parameters $(n, \alpha_0,
\alpha_{\max})$ are $(1.2, 30, 1)$ and the radius is fixed at $ 10.57\,r_g$.
\medskip

\n{\bf Figure 2.} The path of the temperature and the surface density
of model sequence~1 (with initial steady state mass accretion rate
$\mdot=0.3\mdote$) at four specified
radii (in unit of $r_g$): 5.53(a), 11.77(b),
31.10(c) and 59.43(d). The empty circular points represent the initial steady
state model and the cross points correspond to the last evolution model.
The S-shaped curves (dashed lines) are the steady state solutions.
\medskip

\n{\bf Figure 3.} The time variations of the bolometric disk luminosity
in terms of the steady state value of
model sequence~1 with initial steady state mass accretion rate
$\mdot=0.3\mdote$. The disk reachs a stable state after a initial transient
stage of less than 10 seconds.
\medskip

\n{\bf Figure 4.} Same as Figure~2 but for model sequence~2.
The evolution paths of the temperature and the surface density are small
loops at the inner two radii and disappear at the other two larger radii.
\medskip

\n{\bf Figure 5.} The time variations of the disk luminosity
in terms of the steady state value.
(a)--(c) represent model sequences~2--4 respectively.
Note that sequence~2 has a weak periodic oscillations and sequence~4 is stable
due to a lower mass accretion rate.
\medskip

\n{\bf Figure 6.} Same as Figure~4 but for model sequence~5.
In this model, the evolution paths of the temperature and the surface density
at the inner two radii are loops, but they are much larger in comparision
with that of sequence~2, indicating a stronger instability. Note that,
at radius $31.10\,r_g$ (c), a small loop is present even though the initial
model is located at the lower locally stable branch.

\n{\bf Figure 7.} The time variations of the disk luminosity
in terms of the steady state value.
(a)--(c) represent model sequences~5, 6 and 7 respectively.
\medskip

\n{\bf Figure 8.} The time variations of the disk luminosity
in terms of the steady state value.
(a)--(c) represent model sequences~10, 14 and 19 respectively.
\medskip

\n{\bf Figure 9.} The time variations of the radial distribution of the
mass accretion rate of model sequence~19. Note that the variation is global
with the same time scale at different radii, but the unstable region is
restricted to less than $50\,r_g$.

\end{document}